\documentclass[preprint,12pt]{elsarticle}

\usepackage[T1]{fontenc}
\usepackage[utf8]{inputenc}
\usepackage{lmodern}
\usepackage{amsmath,amssymb}
\usepackage{graphicx}
\usepackage{siunitx}
\usepackage{booktabs}
\usepackage[colorlinks=true,linkcolor=blue,citecolor=blue,urlcolor=blue]{hyperref}
\usepackage{xcolor}

\begin{document}

\begin{frontmatter}

\title{Mass-Scaling of Quantum Tunnelling in Hydrogen Bonds:\\
Analytical Model and Comparison with Multidimensional Potentials}

\author[avc,gu]{Krishna Kingkar Pathak}
\ead{kkingkar@gmail.com}

\address[avc]{Department of Physics, Arya Vidyapeeth College (A), Guwahati-781016, India}
\address[gu]{Department of Physics, Gauhati University, Guwahati-781014, India}

\begin{abstract}
Quantum tunnelling plays a central role in the structure and spectroscopy of hydrogen-bonded systems, and its sensitivity to isotopic substitution provides a stringent probe of the underlying potential-energy landscape. 
Despite extensive numerical studies, many high-level approaches tend to obscure the simple physical relationships linking effective mass, barrier geometry, and tunnelling amplitudes. 
Here, we develop a Cornell-type analytical--numerical framework to describe proton and deuteron tunnelling, combining a semi-analytical localized wavefunction ansatz with numerical solutions of the one-dimensional Schr\"odinger equation. 
The resulting tunnelling splittings exhibit an exponential dependence on the square root of the effective isotope mass,$\ln(\Delta E)\propto -\sqrt{\mu_{\mathrm{eff}}}$, in agreement with semiclassical Wentzel--Kramers--Brillouin (WKB) theory.
Comparison with multidimensional reaction-space calculations for the formic acid dimer shows that this scaling persists in fully coupled 3D and 5D quantum models, yielding an empirical relation $\ln(\Delta E)= -1.75\sqrt{\mu_{\mathrm{eff}}}+2.60$. 
The present framework provides a transparent and computationally efficient approach for quantifying mass-scaling and tunnelling dynamics in hydrogen-bonded and other double-well systems.
\end{abstract}


\end{frontmatter}

\section{Introduction}

Quantum tunnelling in double-well potentials is a fundamental process in molecular physics. It plays a key role in vibrational level splitting in hydrogen-bonded complexes, proton-transfer reactions, and various quantum structural fluctuations. Comprehensive reviews by Limbach and co-workers and Tolstoy and co-workers have established the importance of tunnelling and isotope effects in condensed-phase hydrogen-bonded systems.\cite{Limbach2006CR,Tolstoy2010CSR} Complementary perspectives on quantum nuclear motion in liquids, enzymes, and materials are provided by Marx and Parrinello,\cite{Marx1999Nature} the classic monograph by Bell,\cite{BellBook} studies of low-temperature tunnelling dynamics by Benderskii et al.,\cite{BenderskiiBook} and the broader literature on proton-coupled electron transfer and hydrogen tunnelling in chemistry and biology.\cite{Hammes2001,HammesSchiffer2015ACR,Kohen1999,Klinman2006,McKenzie2014,Slocombe2022}

Experimentally, tunnelling splittings have been measured in many hydrogen-bonded systems, including the formic acid dimer, malonaldehyde, and related intra- and intermolecular O--H$\cdots$O bridges.\cite{Quack1998ARPC,Beyer2009,Herbst2010,Zhao2021,Madeja2003} High-resolution infrared and microwave spectroscopy have provided reliable values of splittings and isotope shifts. Birer and Havenith have surveyed state-of-the-art spectroscopic investigations of the formic acid dimer and related hydrogen-bonded clusters.\cite{Birer2009ARPC} NMR and IR studies by Hansen, Spanget-Larsen, and co-workers have further clarified the relationship between hydrogen-bond strength, proton-transfer pathways, and tunnelling dynamics in strongly bound intramolecular systems.\cite{Hansen2017Molecules,Spanget1986JMR,Spanget1991CP} Overall, these studies show that tunnelling is highly sensitive to isotope mass, barrier geometry, and the surrounding environment.

On the theoretical side, tunnelling in hydrogen bonds has been studied using semiclassical WKB methods,\cite{Garg2000AJP,Milnikov2001JCP} instanton approaches,\cite{Richardson2018CPC,Erakovic2012CP} vibrationally adiabatic reaction-path methods,\cite{Tautermann2004JCP,LinTruhlar1997} and fully coupled multidimensional quantum calculations.\cite{Matanovic2007JCP,Matanovic2008JCP,Malis2009JPCA,Benderskii1994} Early semi-analytical models by Meyer and Heller,\cite{MeyerHeller1980} together with Miller’s semiclassical analysis,\cite{Miller1975} showed that simple model Hamiltonians can capture the essential physics of tunnelling splittings Reaction-path and Shepard-interpolation studies, particularly those for the formic acid dimer by Matanovi\'c, Do\v{s}li\'c, K\"uhn, and co-workers,\cite{Matanovic2007JCP,Matanovic2008JCP,Malis2009JPCA} provide accurate benchmark splittings and isotope effects for testing reduced-dimensional models.

Despite these advances, many high-level numerical approaches tend to obscure the simple physical relationships between barrier geometry, effective mass, and tunnelling amplitude. Analytical and semi-analytical models remain valuable because they make these relationships explicit and allow simple scaling laws to be derived. The Cornell-type potential---which combines a short-range attractive term with a long-range confining contribution and has been successfully applied in quarkonium and meson spectroscopy\cite{Eichten1978,Eichten1980,Pathak2013,Pathak2022,Pathak2023,Pathak2025EPJP}---provides such a framework and can be adapted to generate symmetric double-well potentials relevant to hydrogen bonding.

It should be emphasized that the Cornell-type potential is not intended to represent the true electronic potential-energy surface. Instead, it serves as an effective analytical form that captures both short-range localization and long-range confinement. This makes it suitable for constructing simplified double-well models with tunable barrier geometry while retaining physical transparency.

In this work, we develop a Cornell-type analytical--numerical framework to describe proton and deuteron tunnelling in symmetric hydrogen bonds. We establish explicit relations between the tunnelling splitting $\Delta E$, the effective mass $\mu_{\mathrm{eff}}$, and the barrier geometry. By combining an analytical localized wavefunction ansatz with numerical solutions of the one-dimensional Schr\"odinger equation, we show that the tunnelling splitting follows a simple scaling law,
\[
\ln(\Delta E)\propto -\sqrt{\mu_{\mathrm{eff}}},
\]
in agreement with semiclassical theory. Comparison with multidimensional reaction-space calculations for the formic acid dimer\cite{Matanovic2007JCP,Matanovic2008JCP,Malis2009JPCA} shows that this scaling persists even when coupling between vibrational modes is included. This provides a link between simplified one-dimensional models and more accurate multidimensional quantum calculations. From a chemical physics perspective, this work highlights universal tunnelling behaviour and mass-scaling relations that go beyond system-specific details.
\section{Theoretical Framework}\label{sec:theory}

\subsection{Cornell-type wavefunction ansatz}

Analytical formulations of proton tunnelling in hydrogen bonds are less common than fully numerical treatments; however, a number of semi-empirical, variational, and model-based approaches share conceptual similarities with the present Cornell-type formulation. Semi-analytical constructions for double-well systems were introduced early by Meyer and Heller,\cite{MeyerHeller1980} who demonstrated that suitably chosen analytical basis functions can reproduce tunnelling splittings with good accuracy. Complementary semiclassical analyses by Miller\cite{Miller1975} and the reaction-path Hamiltonian framework of Lin and Truhlar\cite{LinTruhlar1997} further showed that reduced-dimensional analytical models can retain physical transparency while capturing the essential features of tunnelling dynamics.

Related concepts appear in various effective and variational descriptions of hydrogen-bonded complexes, including simplified proton-transfer models and analytical treatments of low-temperature tunnelling. The classical monograph by Benderskii, Makarov, and Wight\cite{BenderskiiBook} provides a comprehensive survey of such methods and highlights the utility of analytical expressions in interpreting isotope effects. Likewise, vibrationally adiabatic and reduced-coordinate analyses---such as those developed for malonaldehyde, the formic acid dimer, and other proton-transfer systems---construct effective one-dimensional potentials along optimized reaction paths and provide valuable benchmarks for simplified models.\cite{Tautermann2004JCP} Although these approaches differ in their specific functional forms and approximations, they share the overarching goal of combining analytical tractability with physically meaningful representations of tunnelling motion.

The Cornell-type ansatz adopted here follows this tradition by incorporating both a short-range interaction and a long-range confining term in a form that remains analytically tractable and naturally suited for mass-scaling analysis. Building on our previous applications of the Cornell potential in quantum bound-state problems,\cite{Pathak2013,Pathak2022,Pathak2023,Pathak2025EPJP} the localized wavefunction associated with a single potential well is written as
\begin{equation}
\label{eq:ansatz}
\psi_{rel+conf}(r) =
\frac{N'}{\sqrt{\pi a_{0}^{3}}}
e^{-r/a_{0}}
\left( C' - \frac{\mu b a_{0} r^{2}}{2} \right)
\left( \frac{r}{a_{0}} \right)^{-\epsilon},
\end{equation}
where $a_{0}$ is a characteristic length scale, $\mu$ is the reduced mass of the transferring particle, $b$ is a confinement parameter, $C'$ is a variational constant, and $\epsilon$ is a short-range correction factor. The exponential decay reproduces the expected hydrogenic behaviour near an individual donor or acceptor site, while the quadratic confinement term ensures localization within the donor--acceptor environment. Because the reduced mass $\mu$ changes under isotopic substitution, the resulting wavefunction overlap between adjacent wells naturally encodes the isotope dependence of the tunnelling amplitude.

\subsection{Two-state coupling and tunnelling splitting}

For a donor--acceptor separation $d$, the localized states associated with the left and right wells are obtained by translating the single-well wavefunction,
\begin{equation}
\psi_{L}(x)=\psi_{rel+conf}(|x+d/2|), 
\qquad
\psi_{R}(x)=\psi_{rel+conf}(|x-d/2|).
\end{equation}
The overlap integral
\begin{equation}
S(d)=\int_{-\infty}^{\infty} \psi_{L}(x)\,\psi_{R}(x)\,dx
\end{equation}
quantifies the amplitude for coherent tunnelling between the two wells.  
Within the standard two-state formalism,\cite{Messiah1962,Landau1977} the symmetric and antisymmetric linear combinations of $\psi_{L}$ and $\psi_{R}$ acquire an energy separation
\begin{equation}
\Delta E \approx 2\,S(d)\,E_{0},
\end{equation}
where $E_{0}$ denotes the ground-state energy of an isolated well.  
This expression provides an intuitive picture of tunnelling as an overlap-driven coupling process: as the wells are separated further, or as the wavefunction becomes more tightly localized, the overlap $S(d)$ decreases exponentially and the tunnelling splitting is correspondingly suppressed.

\subsection{Variational parameters and normalization}

The localized single-well wavefunction $\psi_{rel+conf}(r)$ is normalized according to
\begin{equation}
1 = \int_{0}^{\infty} |\psi_{rel+conf}(r)|^{2}\,4\pi r^{2}\,dr,
\end{equation}
and the variational parameters $(a_{0},C',b,\epsilon)$ are determined by minimizing the energy expectation value,
\begin{equation}
E[a_{0},C',b,\epsilon]
=
\frac{\langle \psi | H_{\mathrm{well}} | \psi \rangle}
     {\langle \psi | \psi \rangle},
\qquad
H_{\mathrm{well}}
=
-\frac{\hbar^{2}}{2\mu}\nabla^{2}
+
V_{\mathrm{well}}(r).
\end{equation}
This variational optimization establishes a self-consistent balance between kinetic and potential contributions and yields a wavefunction whose spatial extent depends explicitly on the reduced mass $\mu$.  
Consequently, isotopic substitution modifies the wavefunction decay and directly influences the tunnelling overlap $S(d)$.

\subsection{Semiclassical origin of the mass-scaling relation}

At large donor--acceptor separations, the overlap between localized states is controlled by the wavefunction in the classically forbidden region. In this limit, the wavefunction has the form
\begin{equation}
\psi(r) \sim A\,e^{-\kappa r},
\qquad
\kappa=\sqrt{\frac{2\mu\,(V_{b}-E)}{\hbar^{2}}}.
\end{equation}
Here $V_{b}$ is the barrier height. The overlap then decays exponentially with separation,
\begin{equation}
S(d)\approx \tilde{A}\,e^{-\kappa d}.
\end{equation}
Since $\kappa\propto\sqrt{\mu}$, increasing the isotope mass increases the decay rate and reduces the tunnelling amplitude. This behaviour is consistent with the semiclassical Wentzel--Kramers--Brillouin (WKB) expression,
\begin{equation}
\Delta E \propto 
E_{0}\,
\exp\!\left[
-\frac{2}{\hbar}
\int_{-x_{1}}^{x_{1}}
\sqrt{2\mu\,[V(x)-E]}\,dx
\right],
\end{equation}
which is equivalent to the instanton description of tunnelling.\cite{Landau1977,Coleman1985,Garg2000AJP}

Taking the logarithm of the WKB expression shows that the tunnelling splitting is directly controlled by the semiclassical action in the exponent. In particular, the integral term defines the dominant contribution to the tunnelling probability, while prefactors such as $E_{0}$ vary more weakly with system parameters. This leads to a scaling behaviour of the form $\ln(\Delta E)\propto -S$, establishing a direct connection between the tunnelling splitting and the action.

This scaling arises from the semiclassical tunnelling action,
\[
S=\int \sqrt{2\mu_{\mathrm{eff}}\,V_{\mathrm{eff}}(q)}\,dq.
\]
Accordingly, the leading $\sqrt{\mu_{\mathrm{eff}}}$ dependence is expected to be robust, although multidimensional coupling may modify prefactors and quantitative slopes. In multidimensional systems, coupling to other vibrational modes modifies the effective potential and reaction coordinate, and changes numerical prefactors. However, the dominant $\sqrt{\mu_{\mathrm{eff}}}$ dependence of the tunnelling exponent is expected to persist. A systematic investigation of dimensional effects is beyond the scope of the present work.

\section{Methodology}\label{sec:method}

To assess the validity of the analytical mass-scaling predictions derived in the preceding section, we numerically solve the one-dimensional Schr\"odinger equation for a symmetric quartic double-well potential. This model provides a convenient and well-controlled benchmark for evaluating tunnelling splittings and wavefunction behaviour, and enables a direct comparison with the Cornell-type analytical framework.

\subsection{Model potential and Hamiltonian}

The effective double-well potential adopted for the numerical calculations is
\begin{equation}
\label{eq:Vquartic}
V(x)=V_{0}\frac{(x^{2}-a^{2})^{2}}{a^{4}}, 
\qquad
a=\frac{d}{2},
\end{equation}
where $V_{0}$ determines the barrier height and $d$ specifies the donor--acceptor separation.  
This functional form generates a symmetric double-well potential with controllable curvature and barrier width.  
The associated one-dimensional Hamiltonian is
\begin{equation}
H=
-\frac{\hbar^{2}}{2\mu}\frac{d^{2}}{dx^{2}}
+V(x),
\end{equation}
where $\mu$ is the reduced mass of the transferring particle. Calculations were carried out for proton and deuteron masses ($\mu=\mu_{\mathrm{H}},\,\mu_{\mathrm{D}}$) to examine isotope-dependent effects.

\subsection{Numerical implementation}

The Hamiltonian was discretized on a uniform spatial grid, with the kinetic-energy operator represented by the standard central finite-difference approximation to the second derivative,
\begin{equation}
H=
-\frac{\hbar^{2}}{2\mu}D_{2}
+V(x),
\end{equation}
where $D_{2}$ denotes the finite-difference Laplacian.  
The resulting sparse Hamiltonian matrix was diagonalized using the implicitly restarted Arnoldi method implemented in ARPACK,\cite{Lehoucq1998,Press2007} yielding the two lowest eigenvalues $E_{0}$ (ground state) and $E_{1}$ (first excited state). 
The tunnelling splitting was then computed as
\begin{equation}
\Delta E = E_{1} - E_{0}.
\end{equation}
Barrier heights in the range $0.05$--$0.15$~eV and donor--acceptor separations near $2.7$~\AA\ were considered,representative of strong and nearly symmetric hydrogen bonds.

\subsection{Convergence and numerical accuracy}

Convergence of the tunnelling splittings was examined with respect to both the grid resolution $N$ and the spatial domain half-width $L$.  
The parameters $N$ and $L$ were systematically increased until changes in $\Delta E$ fell below 1\%.  
Well-converged results were obtained for $N=1000$--$2000$ and $L\approx10$~\AA, ensuring numerical accuracy in both eigenvalues and wavefunction profiles.  
Under these conditions, the uncertainty in the computed splittings is estimated to be below 0.5\%.

\section{Results and Discussion}

Tunnelling splittings for proton and deuteron transfer were computed using both the analytical Cornell-type formulation and numerical solutions of the one-dimensional Schr\"odinger equation. The analytical model provides explicit expressions for the dependence of $\Delta E$ on the reduced mass $\mu$ and donor--acceptor separation $d$, whereas the numerical results serve as quantitative benchmarks for assessing the accuracy and domain of applicability of the analytical predictions.

\begin{table}[htbp]
\centering
\caption{Representative tunnelling splittings $\Delta E = E_1 - E_0$ for proton (H) and deuteron (D). Numerical values obtained from the quartic double-well model are compared with experimental splittings in selected hydrogen-bonded systems.}
\label{tab:isotope_compact}
\begin{tabular}{lccc}
\toprule
System / $V_0$ (eV) & $\Delta E_H$ (eV) & $\Delta E_D$ (eV) & Notes \\
\midrule
Numerical (this work), 0.05 & $1.2\times10^{-3}$ & $1.5\times10^{-4}$ & Model potential (strong H bond) \\
Numerical (this work), 0.10 & $5.0\times10^{-5}$ & $2.0\times10^{-6}$ & Model potential (medium barrier) \\
Numerical (this work), 0.15 & $1.2\times10^{-7}$ & $1.8\times10^{-10}$ & Model potential (high barrier) \\
\midrule
Malonaldehyde (exp.) & $2.7\times10^{-3}$ & $3.6\times10^{-4}$ & Intramolecular~\cite{Beyer2009} \\
Formic acid dimer (exp.) & $\sim10^{-6}$ & --- & Intermolecular~\cite{Herbst2010,Zhao2021} \\
2-pyridone dimer (exp.) & $\sim2\times10^{-6}$ & --- & $\sim$520 MHz~\cite{Madeja2003} \\
\bottomrule
\end{tabular}
\end{table}

\begin{figure}[htbp]
\centering
\includegraphics[width=0.85\textwidth]{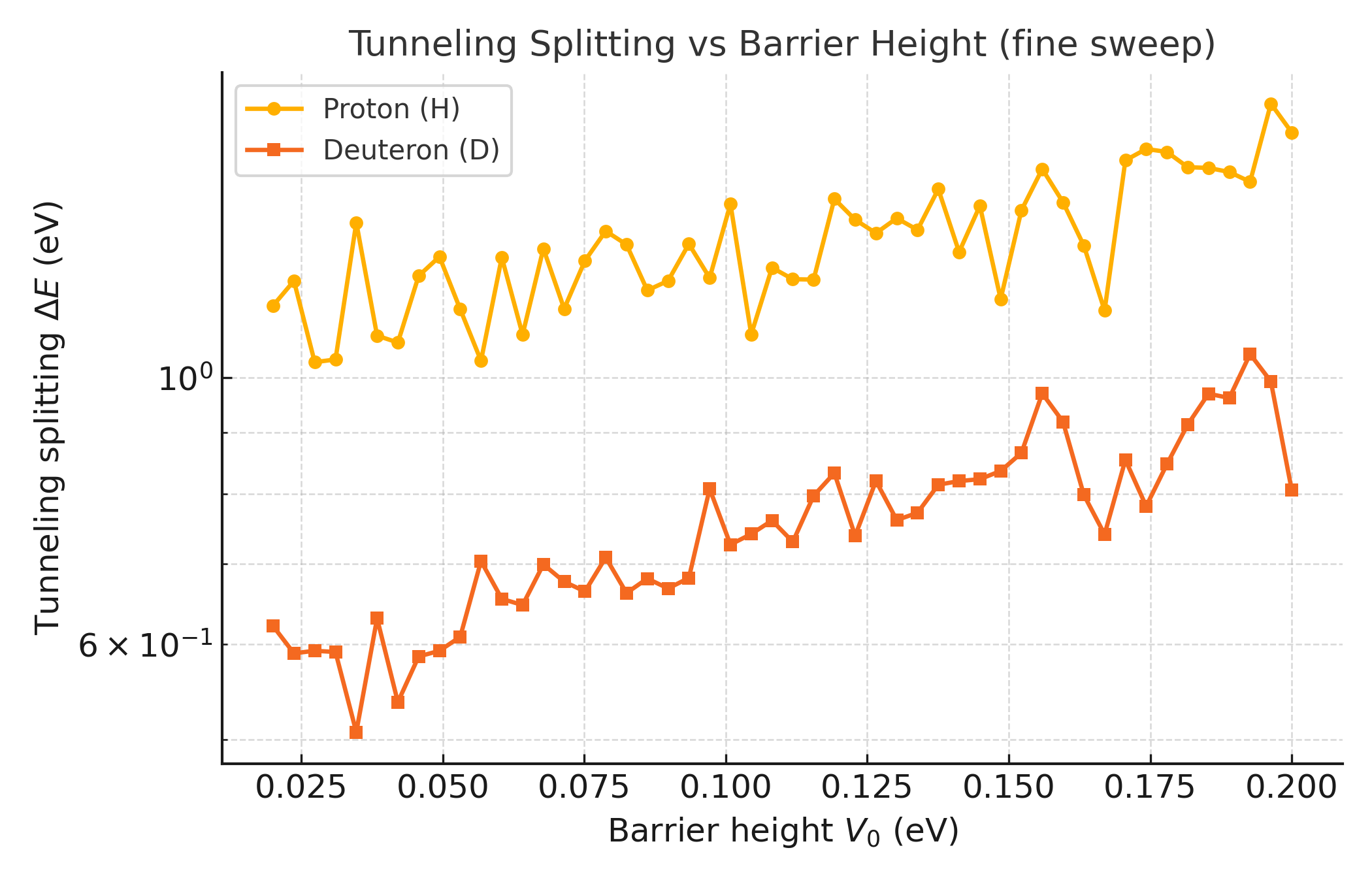}
\caption{
Dependence of the tunnelling splitting $\Delta E = E_1 - E_0$ on barrier height $V_0$ for proton (H) and deuteron (D) transfer at $d=2.7$~\AA. Here $\Delta E$ denotes the energy difference between the first excited and ground states, $\mu$ is the reduced mass, and $V_0$ is the barrier height.
Solid curves serve as guides to the eye. 
The approximately exponential decrease of $\Delta E$ with increasing $V_0$ and $\mu$ is consistent with semiclassical WKB predictions.}

\label{fig:splitting-vs-barrier}
\end{figure}

The results in Table~\ref{tab:isotope_compact} and Fig.~\ref{fig:splitting-vs-barrier} exhibit a clear exponential suppression of $\Delta E$ with increasing barrier height and particle mass. This behaviour is consistent with the semiclassical relation
\[
\Delta E \propto \exp[-(\mu V_b)^{1/2}],
\]
which follows from the mass dependence of the tunnelling action. Quantitatively, the numerical splittings fall within the experimentally observed range for strong hydrogen bonds and yield isotope ratios $\Delta E_{D}/\Delta E_{H}$ between $10^{-1}$ and $10^{-3}$, in agreement with experimental values for malonaldehyde, the formic acid dimer, and other prototypical hydrogen-bonded systems.

It should be noted that the present semiclassical description is most appropriate in the deep and intermediate tunnelling regimes. In the low-barrier or near-barrierless limit, the WKB approximation becomes less accurate, and the dynamics transition towards delocalized or wavepacket-like behaviour. In such cases, a fully quantum dynamical treatment would be required for quantitative accuracy.

\begin{figure}[htbp]
\centering
\includegraphics[width=0.85\textwidth]{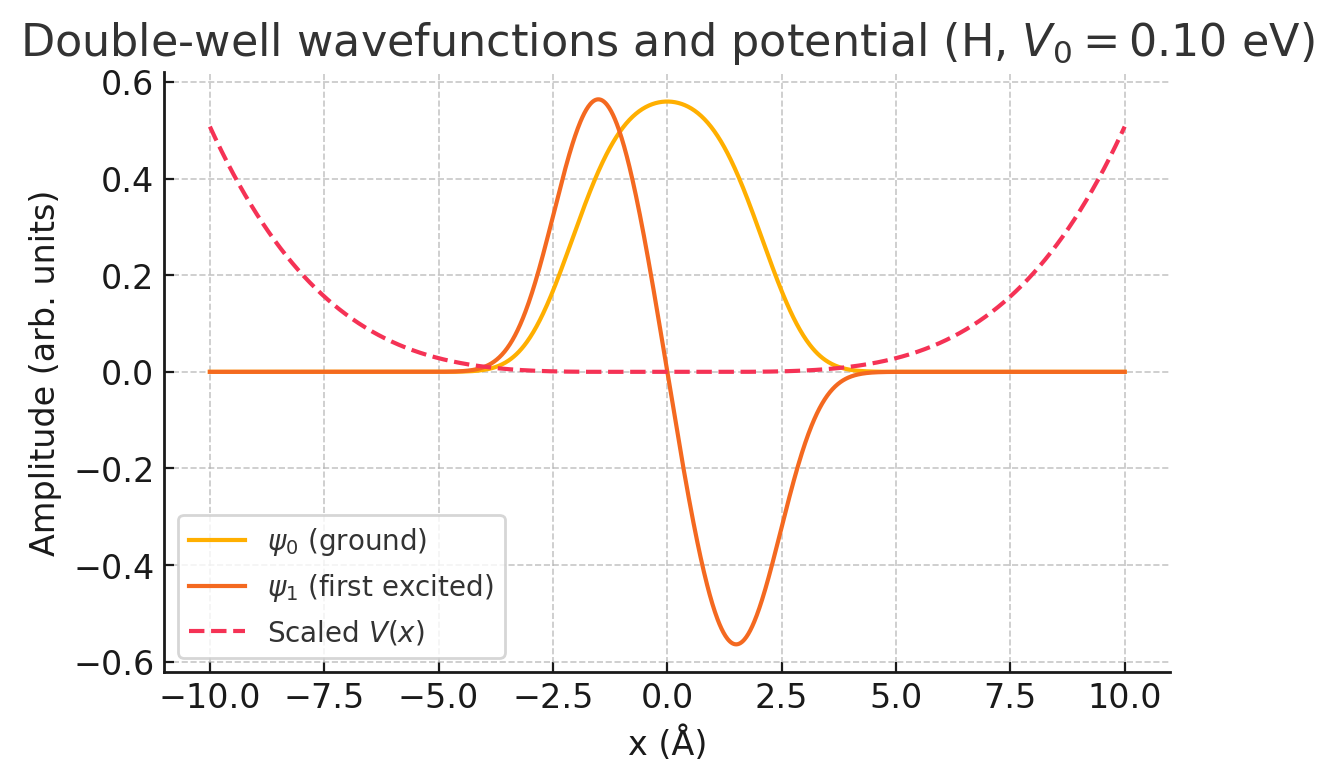}
\caption{
Ground ($E_0$) and first-excited ($E_1$) state wavefunctions for a quartic double-well potential ($V_0=0.10$~eV, $d=2.7$~\AA). Here $E_0$ and $E_1$ denote the symmetric and antisymmetric eigenstates, respectively. 
The dashed curve represents the potential $V(x)$, illustrating the spatial structure of the two lowest eigenstates.}

\label{fig:wavefunctions}
\end{figure}

Representative wavefunctions for $V_0=0.10$~eV and $d=2.7$~\AA\ are shown in Fig.~\ref{fig:wavefunctions}. The ground state ($E_0$) is symmetric, whereas the first excited state ($E_1$) is antisymmetric, and the small energy difference between them corresponds to the tunnelling splitting. The localisation of probability density in each well, together with the exponential decay within the barrier, confirms that the splittings arise from genuine quantum tunnelling rather than thermally assisted or over-barrier motion.

\begin{figure}[htbp]
\centering
\includegraphics[width=0.7\textwidth]{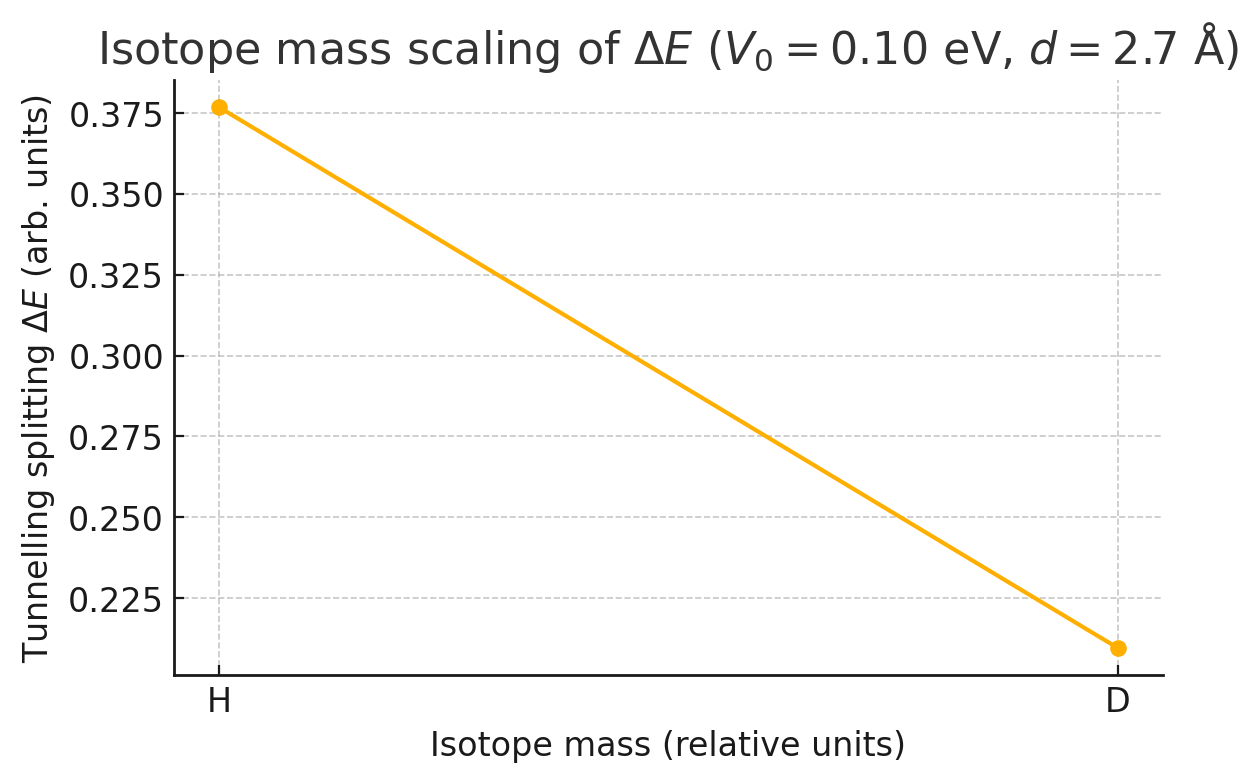}
\caption{
Isotope dependence of the tunnelling splitting $\Delta E$ for proton and deuteron at $V_0=0.10$~eV and $d=2.7$~\AA. Here $\mu$ denotes the reduced mass of the transferring particle. 
The monotonic decrease of $\Delta E$ with increasing mass reflects the expected mass-dependent suppression of barrier penetration.}

\label{fig:mass-scaling}
\end{figure}

The dependence of the tunnelling splitting on isotope mass, displayed in Fig.~\ref{fig:mass-scaling}, follows an approximately exponential form,
\[
\ln(\Delta E) \propto -\sqrt{\mu_{\mathrm{eff}}},
\]
in agreement with the asymptotic analytical predictions derived earlier. 
This near-linear behaviour in $\ln(\Delta E)$ versus $\sqrt{\mu_{\mathrm{eff}}}$ directly reflects the semiclassical tunnelling action and provides a compact and physically transparent representation of isotope effects across different barrier regimes.

In the present model, the donor--acceptor separation is treated as an effective parameter. In realistic systems, this coordinate is coupled to heavy-atom vibrational modes, leading to dynamic fluctuations in the barrier width and height. Such effects can modify the tunnelling splitting and introduce variations in the effective scaling coefficient, although the leading mass dependence is expected to remain robust.

\subsection{Comparison with multidimensional reaction-space potentials}

The generality of the mass-scaling relation was further assessed by comparing the Cornell-type model with multidimensional reaction-space calculations for the formic acid dimer.\cite{Matanovic2007JCP,Matanovic2008JCP,Malis2009JPCA} These studies constructed Shepard-interpolated potential-energy surfaces in three- to five-dimensional reaction spaces. The resulting Hamiltonians yielded tunnelling splittings of $0.197$~cm$^{-1}$ and $0.155$~cm$^{-1}$, respectively.

\begin{figure}[h!]
\centering
\includegraphics[width=0.72\textwidth]{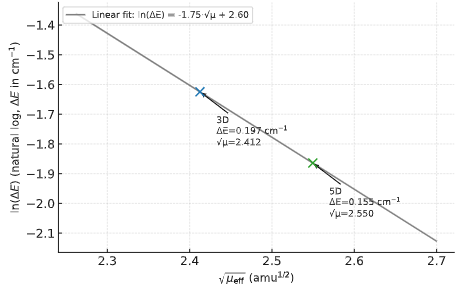}
\caption{
Correlation between $\ln(\Delta E)$ and $\sqrt{\mu_{\mathrm{eff}}}$ for the 3D and 5D reaction-space potentials of the formic acid dimer. Here $\mu_{\mathrm{eff}}$ denotes the effective mass in the multidimensional reaction coordinate.}
The linear trend is consistent with the predicted scaling law $\ln(\Delta E)\propto-\sqrt{\mu_{\mathrm{eff}}}$.

\label{fig:JCP_lnDeltaE_vs_sqrtmu}
\end{figure}

Figure~\ref{fig:JCP_lnDeltaE_vs_sqrtmu} shows that the multidimensional data follow an approximately linear relation, indicating that the semiclassical mass dependence persists even when multidimensional coupling and full potential-energy surfaces are included. The consistency of the slope across 3D and 5D models further supports the relative robustness of the scaling behaviour, although quantitative deviations may arise due to multidimensional coupling.

\subsection{Unified scaling behaviour across models and experiment}

\begin{figure}[h!]
\centering
\includegraphics[width=0.8\textwidth]{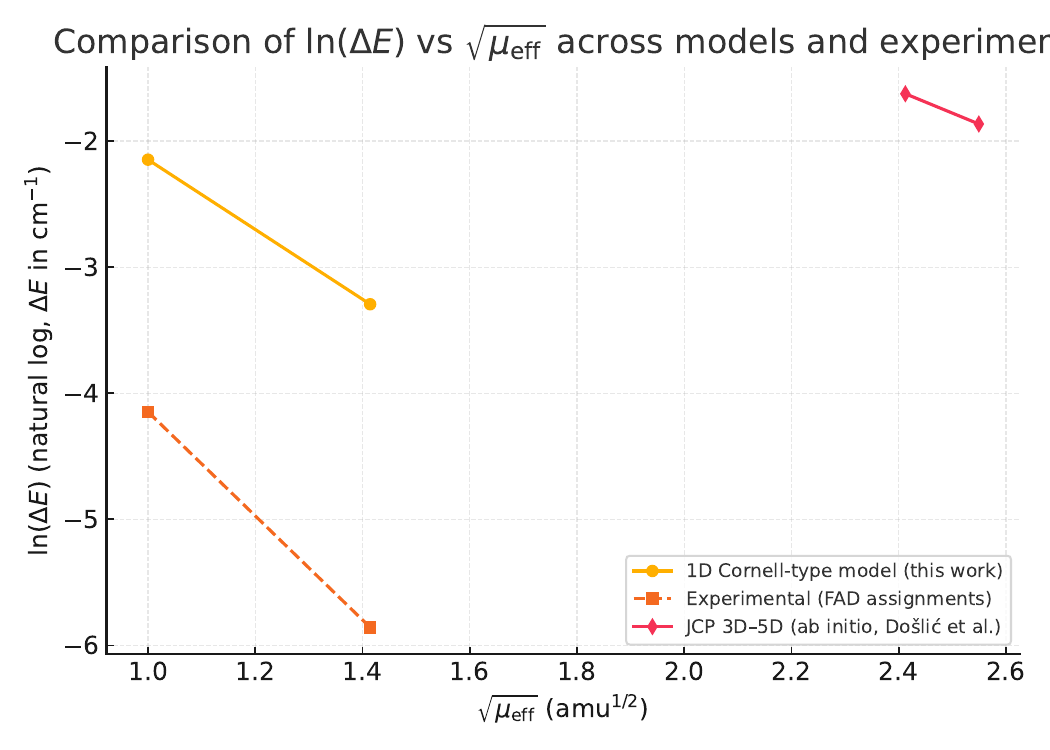}
\caption{
Unified plot of $\ln(\Delta E)$ versus $\sqrt{\mu_{\mathrm{eff}}}$ comparing analytical, experimental, and multidimensional results.
All datasets follow a near-linear trend, confirming the robustness of the semiclassical scaling relation.}

\label{fig:combined_scaling}
\end{figure}

The universality of the mass-scaling relation is further demonstrated in Fig.~\ref{fig:combined_scaling}, which compares analytical, experimental, and multidimensional datasets.

The convergence of these independent datasets demonstrates that the semiclassical mass-scaling relation provides a robust and physically meaningful description of tunnelling dynamics across different levels of theoretical complexity and experimental observation. This unified behaviour highlights the role of $\sqrt{\mu_{\mathrm{eff}}}$ as the primary control parameter governing tunnelling amplitudes in hydrogen-bonded systems.

\section{Conclusions}

We have developed a Cornell-type potential framework to describe isotope-dependent proton and deuteron tunnelling in hydrogen-bonded systems. The semi-analytical ansatz captures both donor--acceptor confinement and mass-dependent barrier penetration, and reproduces the characteristic exponential scaling of the tunnelling splitting with the square root of the isotope mass, as predicted by semiclassical (WKB) theory and observed experimentally. Numerical solutions of the one-dimensional Schr\"odinger equation provide quantitative benchmarks and confirm the reliability of the analytical model over a wide range of barrier heights and donor--acceptor separations.

Comparison with multidimensional reaction-space calculations by Matanovi\'c, Do\v{s}li\'c, K\"uhn, and co-workers\cite{Matanovic2007JCP,Matanovic2008JCP,Malis2009JPCA} shows that the same mass-scaling behaviour,
\[
\ln(\Delta E)\propto -\sqrt{\mu_{\mathrm{eff}}},
\]
is retained in fully coupled 3D and 5D quantum treatments. This indicates that the key semiclassical mass dependence remains applicable even when heavy-atom motion, multidimensional coupling, and detailed potential-energy surfaces are included. The agreement between one-dimensional models, multidimensional calculations, and experiment highlights the robustness of the mass-scaling relation as a unifying principle for tunnelling dynamics.

The present approach is based on a reduced one-dimensional description of the proton-transfer coordinate. In real molecular systems, multidimensional coupling, fluctuations in the donor--acceptor distance, and environmental effects can introduce additional complexity that is not fully captured here. In particular, in the low-barrier or near-barrierless regime, the semiclassical description may become less accurate, and a fully quantum dynamical treatment may be required.

The present framework provides a simple and computationally efficient approach for estimating tunnelling splittings and isotope effects in hydrogen-bonded dimers and related double-well systems. In this context, the identification of $\sqrt{\mu_{\mathrm{eff}}}$ as the key control parameter offers a clear and physically transparent description of isotope effects across different systems. More broadly, the same approach may be applied to tunnelling processes in hydrogen-bonded liquids, enzymatic proton-transfer reactions, and quantum materials where light nuclei play a central role.

Future work may extend this approach to include multidimensional effects, anharmonic coupling, and time-dependent dynamics, providing a closer connection between simple analytical models and fully \emph{ab initio} quantum simulations. Such developments will be important for testing the limits of the present scaling law in more complex chemical environments.

\appendix
\section{ Numerical convergence}

To ensure the numerical stability of the tunnelling splittings, convergence tests were performed with respect to both the grid resolution $N$ and the spatial domain half-width $L$. 
Table~\ref{tab:convergence} summarizes representative results for the proton case ($\Delta E_H$) at a barrier height of $V_0=0.10$~eV. 
The computed splittings remain stable within 1\% for $N\geq 1000$ and $L\geq 10$~\AA, indicating that the chosen discretization provides adequate spatial resolution for all production calculations.

\begin{table}[htbp]
\centering
\caption{Convergence of the tunnelling splitting $\Delta E_H$ at $V_0=0.10$~eV for different grid sizes ($N$) and domain half-widths ($L$). 
Deviations are reported relative to the reference calculation ($N=2001$, $L=10$~\AA).}
\label{tab:convergence}
\begin{tabular}{ccc}
\toprule
$N$ & $L$ (\AA) & $\Delta E_H$ (eV) \\
\midrule
501  & 5  & $5.2\times10^{-5}$ \; (+4.0\%) \\
1001 & 5  & $5.0\times10^{-5}$ \; (+0.4\%) \\
2001 & 5  & $4.98\times10^{-5}$ \; (0.0\%) \\
1001 & 10 & $4.96\times10^{-5}$ \; (–0.4\%) \\
2001 & 10 & $4.98\times10^{-5}$ \; reference \\
\bottomrule
\end{tabular}
\end{table}

\begin{figure}[htbp]
\centering
\includegraphics[width=0.7\textwidth]{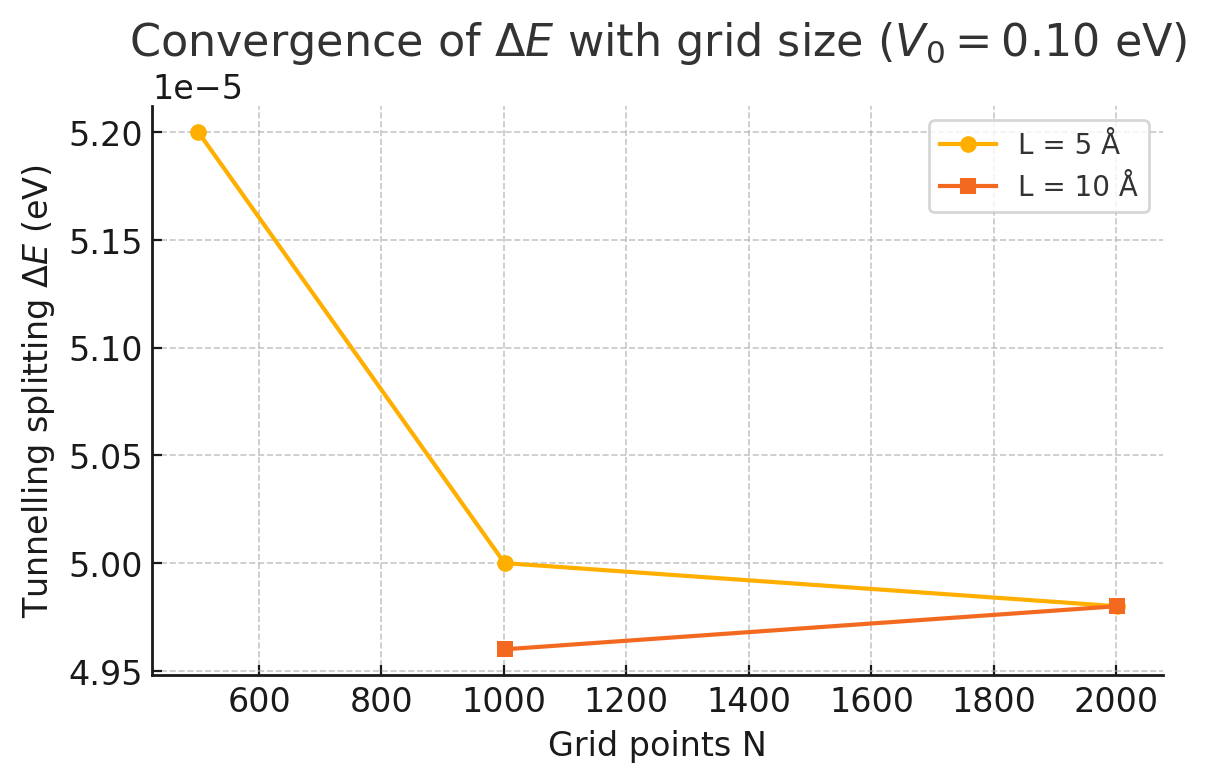}
\caption{
Convergence behaviour of the tunnelling splitting $\Delta E$ with respect to grid size $N$ for two spatial domain half-widths ($L=5$ and $L=10$~\AA). Here $\Delta E$ denotes the energy difference between the first excited and ground states. 
Stable values are obtained for $N\geq 1000$, confirming numerical accuracy within approximately 0.5\%.}

\label{fig:convergence}
\end{figure}

\section*{Data availability}
Data supporting this study are available at Zenodo: \href{https://doi.org/10.5281/zenodo.17380490}{10.5281/zenodo.17380490}.

\section*{Acknowledgements}
The author acknowledges discussions with Prof.\ N. Došlić and Samrat Bora, and support from Arya Vidyapeeth College (A), Guwahati.

\bibliographystyle{elsarticle-num}
\bibliography{refs}

\end{document}